\documentclass[a4paper]{jpconf}
\usepackage{graphicx}

\newcommand {\zyam}	{{\sc zyam}}
\newcommand {\gev}	{{GeV/$c$}}
\newcommand {\dAu}	{{$d$+Au}}
\newcommand {\etl}	{{\it et al.}}
\newcommand {\mn}[1]	{\langle#1\rangle}
\newcommand {\psiEP}	{\psi_{\rm EP}}
\newcommand {\ppt}	{p_{T}}
\newcommand {\ptt}	{p_{T}^{(t)}}
\newcommand {\pta}	{p_{T}^{(a)}}
\newcommand {\phis}	{\phi_{s}}
\newcommand {\phit}	{\phi_{t}}
\newcommand {\dphi}	{\Delta\phi}
\newcommand {\deta}	{\Delta\eta}
\newcommand {\vvvt}	{v_3^{(t)}}
\newcommand {\va}	{v_2^{(a)}}
\newcommand {\vvva}	{v_3^{(a)}}
\newcommand {\vtR}	{v_2^{(t,R)}}
\newcommand {\vvaPsi}	{v_4^{(a)}\{\psi_2\}}
\newcommand {\vvtRPsi}	{v_4^{(t,R)}\{\psi_2\}}
\newcommand {\ff}[2]	{v_{#1}\{#2\}}
\newcommand {\fff}[3]	{v_{#1}^{(#2)}\{#3\}}
\newcommand {\etagap}	{\eta_{\rm gap}}
\newcommand {\VVuc}     {V_4\{{\rm uc}\}}
\newcommand {\veta}[2]	{v_{#1}\{2,\etagap\mbox{=}#2\}}

\newcommand {\vpt}[1]   {v_{#1}\{\ppt\mbox{-}\ppt\}}
\newcommand {\vpteta}[2]{v_{#1}\{\ppt\mbox{-}\ppt,\etagap\mbox{=}#2\}}
\newcommand {\Vpteta}[2]{V_{#1}\{\ppt\mbox{-}\ppt,\etagap\mbox{=}#2\}}

\begin{document}
\title{Dihadron Correlations Relative to the Event Plane in 200 GeV Au+Au Collisions from STAR}
\author{Fuqiang Wang (for the STAR Collaboration)}
\address{Department of Physics, Purdue University, West Lafayette, Indiana 47907, USA}
\ead{fqwang@purdue.edu}
\begin{abstract}
Dihadron correlations with a high-$\ppt$ trigger particle are analyzed by STAR relative to the event plane in Au+Au collisions at 200 GeV (arXiv:1010.0690v1). The elliptic and quadrangular flow anisotropies are subtracted. The remaining dihadron correlation signals are found to be composed of a near-side peak, which is separated into jet-like and ``ridge''-like components, and an away-side correlation structure. The ridge-like structure is found to decrease with the trigger particle azimuthal angle relative to the event plane from in-plane to out-of-plane. The away-side structure is found to evolve from single-peak for in-plane triggers to double-peak for out-of-plane triggers. Is the dihadron correlation signal just a manifestation of the triangular and higher-order harmonic flows? This talk addresses this question, exploring the answers using limiting-case assumptions, and discusses the implications of the data regarding the ridge and the double-peak structure.
\end{abstract}

\section{Introduction}

Collisions at the Relativistic Heavy Ion Collider (RHIC) of Brookhaven National Laboratory have created a hot and dense medium with properties that resemble a nearly perfect fluid of strongly interacting quarks and gluons (sQGP)~\cite{wpBRAHMS,wpPHOBOS,wpSTAR,wpPHENIX}. Energetic partons produced by rare hard-scattering processes traverse and interact with the sQGP, losing energies resulting in suppression of high transverse momentum ($\ppt$) particle yields and modifications to jet-like correlations~\cite{b2b,jetspec}. Several novel phenomena have been observed in jet-like correlations with high-$\ppt$ trigger particles. On the near-side of the high-$\ppt$ trigger particle, correlated particles in $\dphi$ are found to extend to large pseudo-rapidities ($\eta$)~\cite{jetspec,ridge,Pawan,PHOBOS}. On the away-side of the trigger particle, the correlation function is found to be double-peaked away from the back-to-back direction of $\dphi=\pi$~\cite{jetspec,Lacey,Horner,Jia}. 

The effect of jet-medium interactions is expected to depend on the path-length of the sQGP medium the away-side jet traverses. STAR has analyzed dihadron correlations as a function of the trigger particle relative to the event plane in arXiv:1010.0690v1~\cite{STAR}. Indeed the correlations are found to be dependent on the trigger particle direction. In particular, the ridge correlation magnitude is strongest for in-plane trigger particles and decreases with increasing $\phis$ from in-plane to out-of-plane. The away-side correlation is found to be single peaked at $\dphi=\pi$ for in-plane triggers, but become double-peaked for out-of-plane triggers. The results indicate that the ridge is dominated by in-plane triggers and the away-side double-peak is dominated by out-of-plane triggers.

The major backgrounds to jet-like correlations are two-particle correlations caused by single-particle anisotropic flows. The elliptic and quadrangular flow backgrounds have been subtracted in dihadron correlations as a function of $\phis$ in~\cite{STAR}. It has been recently suggested, however, that there should be triangular anisotropic flow~\cite{Mishra,Alver}. The triangular anisotropy is a result of hydrodynamics evolution from initial-state geometry triangularity due to event-by-event fluctuations~\cite{Petersen,Schenke,Heinz}. It is demonstrated illustratively by the NeXSPheRIO model~\cite{Takahashi,Hama,Andrade} that initial energy density fluctuations (hot spots) with subsequent hydro evolution may generate a near-side ridge and a double-peak correlation on the away side. 
Triangular anisotropic flow would result in three peaks at $\dphi=0$, $2\pi/3$, and $4\pi/3$ in two-particle correlations. These peaks would qualitatively explain the observed ridge and double-peak structure in dihadron correlations. 

The natural question is then, whether hydrodynamic triangular flow can explain entirely the observed ridge and double-peak correlations? This talk attempts to answer this question.

The current understanding of anisotropic flow backgrounds, up to the fourth-order harmonic, is given by:
\begin{eqnarray}
\frac{dN}{d\dphi}&=&B\left[1+2\va\vtR\cos(2\dphi)+2\vvaPsi\vvtRPsi\cos(4\dphi)\right.\nonumber\\
&& \hspace{5mm}\left. + 2\vvva\vvvt\cos(3\dphi)+2\VVuc\cos(4\dphi)\right].\label{eq}
\end{eqnarray}
Here $v_n^{(t)}$ and $v_n^{(a)}$ are the trigger and associated particle $n^{th}$ harmonic flow anisotropy, respectively; $v_2^{(t,R)}=\mn{\cos2(\phit-\psi_2)}^{(R)}$ and $v_4^{(t,R)}=\mn{\cos4(\phit-\psi_2)}^{(R)}$ are, respectively, the trigger particle second and fourth harmonic anisotropies with respect to the second harmonic plane, $\psi_2$ and averaged with a given $\phis$ range (see~\cite{STAR} for details); and $\VVuc$ is the component of the fourth harmonic anisotropy uncorrelated to $\psi_2$. It is assumed in Eq.~(\ref{eq}), and verified by experiments~\cite{PHENIX_v3,STAR_v3}, that $v_3$ fluctuation effects are uncorrelated to $\psi_2$. 

In the following sections we discuss the flow backgrounds of Eq.~(\ref{eq}) in detail. We focus on the dihadron correlations at large $|\deta|>0.7$ in the interest to study the near-side ridge correlations as well as the away-side correlations.

\section{Systematic uncertainty of $v_2$}
\begin{figure}[hbt]
  \centerline{\includegraphics[width=1.03\textwidth]{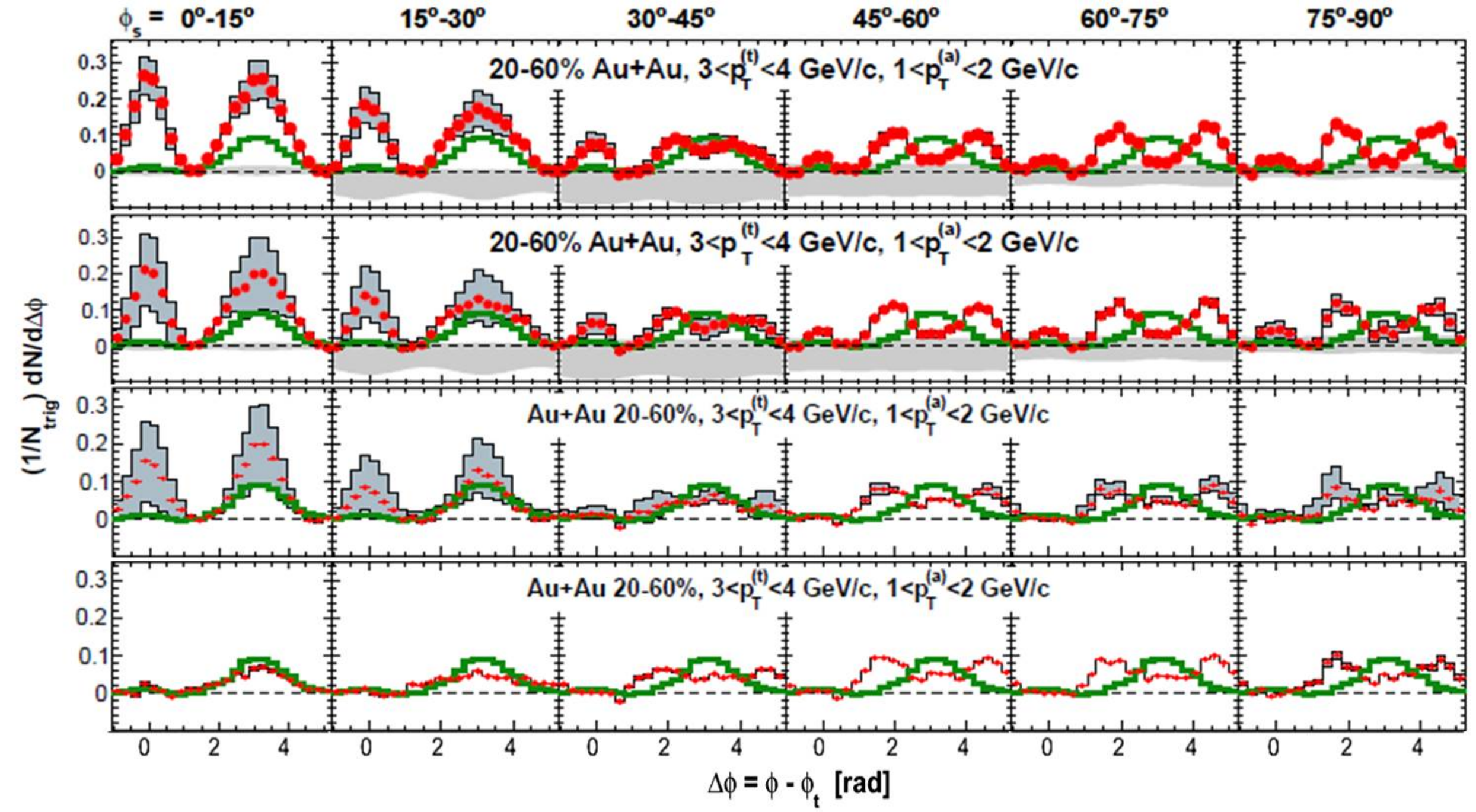}}
  \caption{(Color online) Anisotropic background subtracted dihadron correlations with trigger particles in six slices of azimuthal angle relative to the event plane, $\phis=|\phit-\psiEP|$, with a cut on the trigger-associated pseudo-rapidity difference of $|\deta|>0.7$. The triangle two-particle $\deta$ acceptance is not corrected. The trigger and associated particle $\ppt$ ranges are $3<\ptt<4$~\gev\ and $1<\pta<2$~\gev, respectively. The data points are from minimum-bias 20-60\% Au+Au collisions. The backgrounds due to elliptic anisotropy $v_2$ and quadrangular anisotropy relative to the second harmonic plane, $\ff{4}{\psi_2}=1.15v_2^2$ are subtracted. Shown in the thin histograms embracing the shaded area are systematic uncertainties due to flow subtraction, the two- and four-particle cumulant $v_2$ (with the average used for the data points). The systematic uncertainty due to \zyam\ normalization are shown as the horizontal shaded band around zero in the top two rows. For comparison, the inclusive dihadron correlations from \dAu\ collisions are superimposed as the thick (green) histograms.
First row: the upper systematic uncertainty of $v_2$ is $\mn{\cos2\dphi}$ average over the away-side of two-particle cumulant.
Second row: the upper systematic uncertainty of $v_2$ is $\mn{\cos2\dphi}$ average over the entire $2\pi$ range. 
Third row: an additional $v_3$ background, obtained from two-particle cumulant with $\eta$-gap of $\etagap=0.7$, is subtracted.
Fourth row: The $\phis$-dependent $\vpt{2}$ measured by two-particle cumulants with $\etagap=0.7$ and 1.2 are used (the thin histograms embracing the shaded area), with their average shown in the data points. An additional, but small $\psi_2$-uncorrelated $\VVuc$ is subtracted. The subtracted $v_3$ is as same as the third row. The data should be compared to the lower systematic bounds in the third row where the subtracted $v_2$ is the $\phis$-independent two-particle cumulant measurement.}
  \label{fig}
\end{figure}

For the dihadron correlation data presented in~\cite{STAR}, the maximum systematic uncertainty bound of $v_2$ is obtained by $\mn{\cos2\dphi}$ averaged over the away-side of two-particle correlations between the particle of interest and a reference particle in the flow measurements in order to avoid near-side small angle nonflow contributions. The results~\cite{STAR} are shown in the first row of Fig.~\ref{fig}. This would be the correct maximum systematic bound of $v_2$ if only even harmonic flows are present. With non-vanishing odd harmonic flows, this estimate of the maximum systematic bound would not necessarily be the maximum $v_2$. Therefore, we now use the $\mn{\cos2\dphi}$ averaged over the entire $2\pi$ range as the upper bound of the $v_2$ systematic uncertainty. This is larger than the previous estimate because of the inclusion of large nonflow contributions from near-side correlations. The two-particle correlations to obtain the maximum $v_2$ are restricted to $|\deta|>0.7$ because the dihadron correlations we study are from the same $|\deta|>0.7$ region.
The new estimated systematic upper bound is the maximum possible $v_2$ that should be subtracted from raw dihadron correlations. 
The low systematic bound of $v_2$ is still the four-particle cumulant measurement as in~\cite{STAR}. The default $v_2$ is the average of $\ff{2}{4}$ and $\ff{2}{2}$. The dihadron correlation results with the new estimate of $v_2$ systematic uncertainties are shown in the second row of Fig.~\ref{fig}. 

\section{Subtraction of $v_3$}
The $v_3$ of trigger and associated particles are measured by the two-particle cumulant method with a reference particle of $0.15<\ppt<2$~\gev~\cite{STAR_v3}. A $\eta$-gap ($\etagap$) of 0.7 is applied between the particle of interest and the reference particle. 

The third row of Fig.~\ref{fig} shows the dihadron correlation functions for $|\deta|>0.7$ obtained with $\ff{3}{2}$ included in the background subtraction. The change from the second row is the additional subtraction of the $\ff{3}{2}$ contribution. The qualitative features of the correlation functions are unchanged. The away-side double-peak structure out-of-plane remains prominent. The decreasing trend of the ridge magnitude from in-plane to out-of-plane is unaffected because a constant $v_3$ contribution over $\phis$ is subtracted. This indicates that the main features of the measured near-side ridge and away-side double-peak in the dihadron correlations with high $\ppt$ trigger particles, whether or not integrated over $\phis$, are not entirely due to the triangular flow contributions, but other physics mechanisms.

\section{Subtraction of uncorrelated $v_4$\label{sec:v4uc}}
We have so far subtracted the $\ff{4}{\psi_2}$ background correlated with the second harmonic plane, $\psi_2$. We have used the parameterization to the previous $\ff{4}{\psi_2}$ measurement~\cite{v2MRP},
\begin{equation}
\ff{4}{\psi_2}=1.15v_2^2\,.
\end{equation}

There is an additional contribution to $v_4$ that is uncorrelated to $\psi_2$ and arises from fluctuations. The uncorrelated component can be obtained by
\begin{equation}
\VVuc=\fff{4}{t}{2}\fff{4}{a}{2}-\fff{4}{t}{\psi_2}\fff{4}{a}{\psi_2}\,.\label{eq:v4uc}
\end{equation}
where $\ff{4}{2}$ is the $v_4$ measured by the two-particle cumulant method with $\etagap=0.7$.
Because $\VVuc$ is small, its effect on dihadron correlation is negligible. The dihadron correlation results with $\VVuc$ subtraction is effectively as same as those shown in the third row of Fig.~\ref{fig}.

\section{Effect of possible $\phis$-dependent $v_2$}

One can always attribute all azimuthal dependence to Fourier harmonics. In fact, Luzum~\cite{Luzum} argued that our $|\deta|>0.7$ correlation data can be fitted by Fourier harmonics up to the $4^{\rm th}$ order and the fitted coefficients are consistent with features expected from anisotropic flows. This is not surprising because nonflow effects, which must be contained in the fitted Fourier coefficients, are relatively small compared to the flow contributions in our kinematic regions. If the observed $\phis$-dependent ridge is due to anisotropic flow, then the harmonic flows must be $\phis$-dependent. This may not be impossible, however, as the requirement of trigger particles in a particular $\phis$ bin from the event plane reconstructed from particles in $0.15<\ppt<2$~\gev\ could preferentially select events with associated particle $v_2$ displaced from the average. In the following, we analyze the two-particle cumulant $v_n$ in events of different $\phis$ values separately, and subtract them from the dihadron correlations. 

\begin{figure}[hbt]
\centerline{\includegraphics[width=0.45\textwidth]{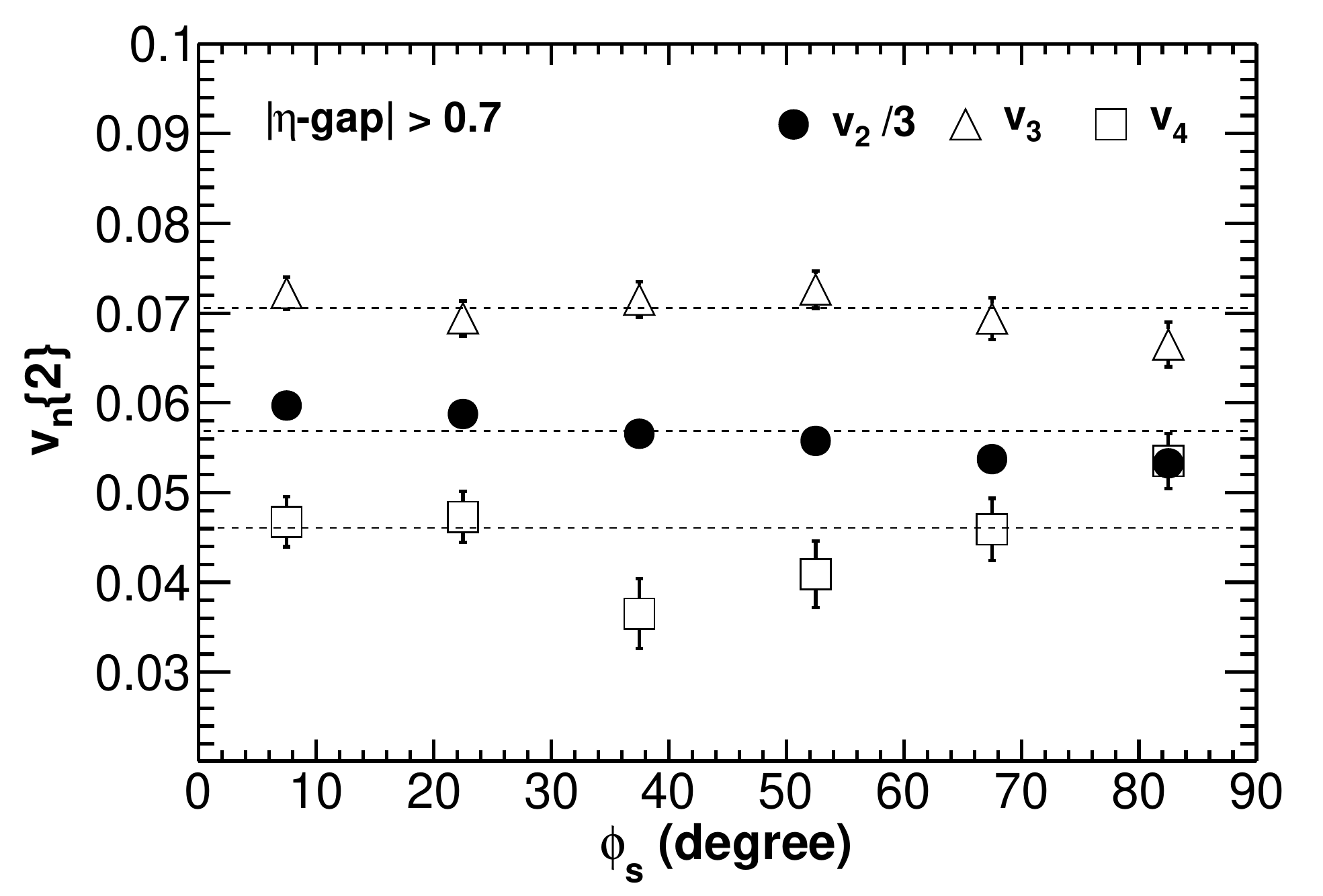}}
\caption{Harmonic $\vpt{n}$ of associated particles of $1.5<\pta<2$~\gev\ as a function of $\phis$ of trigger particles of $3<\ptt<4$~\gev. Note the $\vpt{2}$ is scaled down by a factor of 3 to fit into the plot coordinate range. The $\vpt{n}$ is measured by the two-particle cumulant method with particle pairs from the same associated $\pta$ bin and with $\etagap=0.7$. The data are from minimum-bias 20-60\% Au+Au collisions. Error bars are statistical. The horizontal lines are to guide the eye.}
\label{fig:vn_phisDep}
\end{figure}

Since reference particles are used to reconstruct the event plane to determine the $\phis$, one cannot calculate $v_n$ from cumulant of the associated particle and a reference particle in event sample selected according to $\phis$. Instead, we form two-particle cumulant from particles in a given associated $\pta$ bin, applying $\eta$-gap of 0.7. The $v_n$ of the associated particles are simply 
\begin{equation}
\vpt{n}(\phis)=\sqrt{\Vpteta{n}{0.7}(\phis)}\,.\label{eq:vn_phisDep}
\end{equation}
Here $\Vpteta{n}{0.7}$ indicates the two-particle cumulant with particle pairs from the same $\ppt$ bin. We use $\vpteta{n}{0.7}$ or simply $\vpt{n}$ to stand for the resultant anisotropy measurement. Figure~\ref{fig:vn_phisDep} shows the obtained $\vpt{n}$ of $1.5<\ppt<2$~\gev\ as a function of $\phis$ of trigger particles of $3<\ptt<4$~\gev. The $\vpt{2}$ decreases with $\phis$. The decrease is a consequence of the decreasing ridge with increasing $\phis$. The $\vpt{4}$ is found to be smallest with $\phis=45^\circ$ and largest with $\phis=0^\circ$ and $90^\circ$. On the other hand, the $\vpt{3}$ is independent of $\phis$, consistent with the expectation that the third and second harmonic planes are uncorrelated. The $\vpt{3}$ from the cumulant of same-$\ppt$ bin pairs is consistent with that obtained from the cumulant with a reference particle, $\ff{3}{2}$. 

Although the measured $\vpt{4}$ is $\phis$-dependent, the contribution of the $\psi_2$-uncorrelated $v_4$ to flow background is negligibly small, as discussed in Sec.~\ref{sec:v4uc}. We therefore use the $\phis$-independent $\ff{4}{2}$ measured by two-particle cumulant with a reference particle, as in Sec.~\ref{sec:v4uc}. We have checked our results using the $\phis$-dependent $\vpt{4}$ and found no observable difference. 

We subtract flow background using the $\phis$-dependent $\vpt{2}(\phis)$. The trigger particle $v_2$ is still given by the two-particle cumulant flow obtained with a reference particle. This is because the trigger $v_2$ is the second harmonic modulation of trigger particles which determines the $\phis$. 
The bottom row of Fig.~\ref{fig} shows the dihadron correlation results with subtraction of $\phis$-dependent $\vpt{2}$. The change from the lower systematic bound in the third row is the subtraction of the $\phis$-dependent $v_2$ in place of the $\phis$-independent one.

As seen from figure, the near-side ridge is diminished, maybe as expected, because the large $\deta$ ridge is presumably included in the subtracted $v_n$. However, it is important to point out that it is not automatically guaranteed that the ridge will be gone just because the $v_n$'s are measured by two-particle cumulant either with a reference particle or with a particle from the same $\ppt$ region. This is because they are not simply measured by the trigger-associated particle pair at $|\deta|>0.7$. If they were, then the correlation will be strictly zero everywhere, both on the near side and on the away side. 

It is interesting to note, despite of the diminished near-side ridge, that the away-side correlation is not diminished. It still evolves from a single peak with in-plane trigger particles to a double peak with out-of-plane trigger particles. The observation of the away-side double-peak structure for out-of-plane triggers seems robust against the wide range of flow background subtraction.

Since the ridge is diminished after subtraction of $\phis$-dependent $v_2$ from two-particle cumulant at large $\deta$, can we conclude that the physics origin of the ridge is hydrodynamic $v_n$ flow? The answer is no because any non-hydrodynamic origin of $v_n$ is also included in the two-particle $v_n$ measurements. In other words, any ridge signal (whatever its physics origin might be) is included in $v_n$, and the ridge would be subtracted after subtraction of $v_n$. However, one also cannot rule out the ridge being part of hydrodynamic flow. This is because it is still possible that hydrodynamic flow of the underlying event is biased by the selection of the trigger particle orientation, and all the long-range $\deta$ correlation may be indeed due to flow. 

\subsection{Implications on inclusive dihadron correlations}
If $v_2$ depends on $\phis$, then there is an important implication to the inclusive dihadron correlation (i.e., without cutting on $\phis$). For inclusive dihadron correlation, a flow background $\mn{\fff{2}{t}{2}}\cdot\mn{\fff{2}{a}{2}}$ has been used in place of $\mn{\fff{2}{t}{2}\cdot\fff{2}{a}{2}}$. (Note, for clarity, we have omitted the $\mn{...}$ notation throughout the article except here.) This is correct because fluctuations are already included in the two-particle cumulant flow measurement of $\mn{\ff{2}{2}}$. However, if $v_2$ depends on trigger particle orientation $\phis$, then the equality $\mn{\fff{2}{t}{2}(\phis)\cdot\fff{2}{a}{2}(\phis)}=\mn{\fff{2}{t}{2}(\phis)}\cdot\mn{\fff{2}{a}{2}(\phis)}$ is no longer valid. The left side will be always larger than the right side. This means that the inclusive dihadron flow background is underestimated by $\mn{\fff{2}{t}{2}}\cdot\mn{\fff{2}{a}{2}}$. In fact, because $\fff{2}{t}{2}(\phis)$ is positive for $\phis\sim0$ and negative for $\phis\sim\pi/2$, the true background magnitude for inclusive dihadron correlation is even larger than that for the $\phis=0$ dihadron correlation, which has the largest background magnitude of all $\phis$ bins. Namely
$\mn{\fff{2}{t}{2}(\phis)\cdot\fff{2}{a}{2}(\phis)} > \mn{\fff{2}{t}{2}}\cdot\fff{2}{a}{2}(\phis=0)$.

\begin{figure}[hbt]
  \begin{center}
    \includegraphics[width=0.35\textwidth]{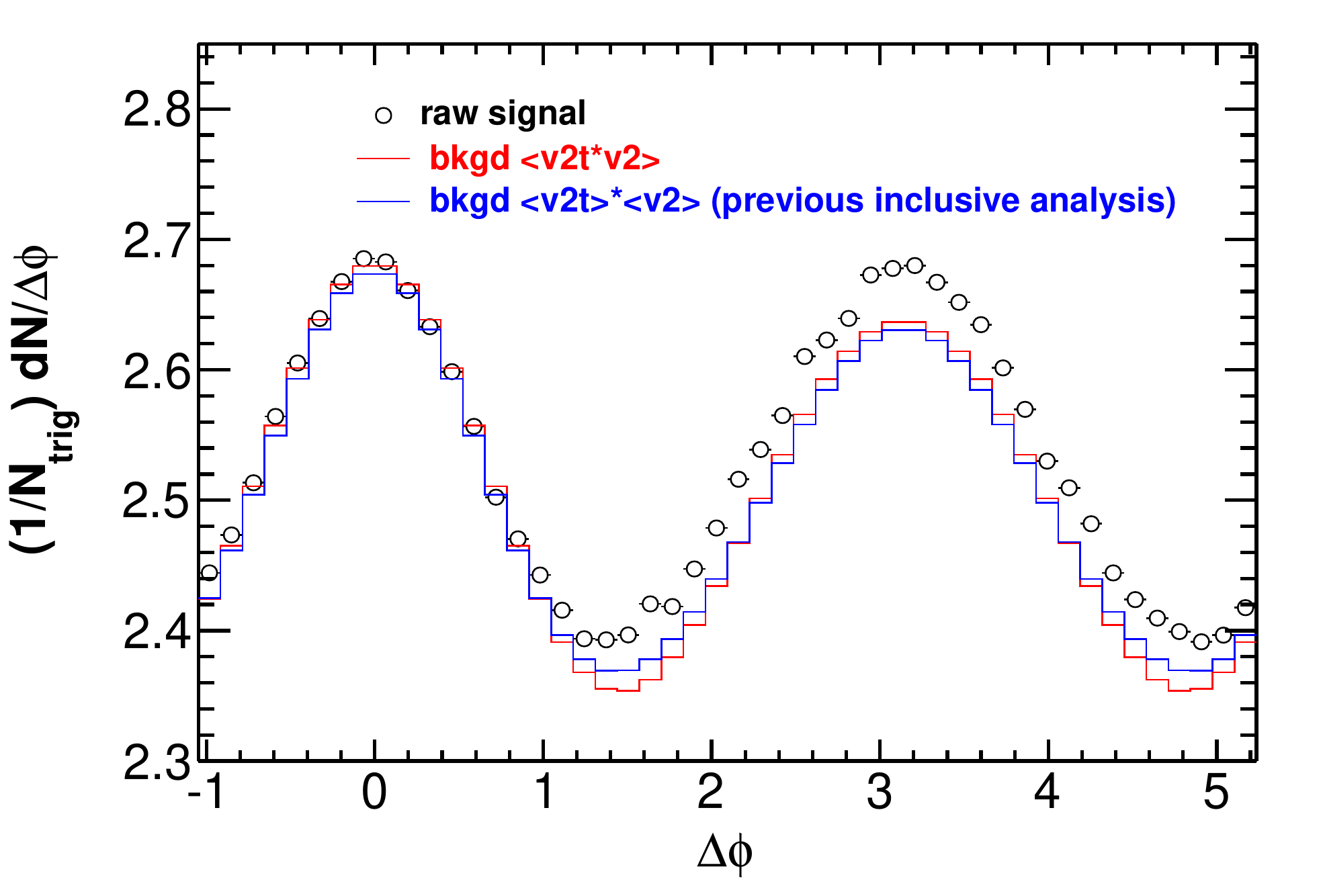}
    \includegraphics[width=0.35\textwidth]{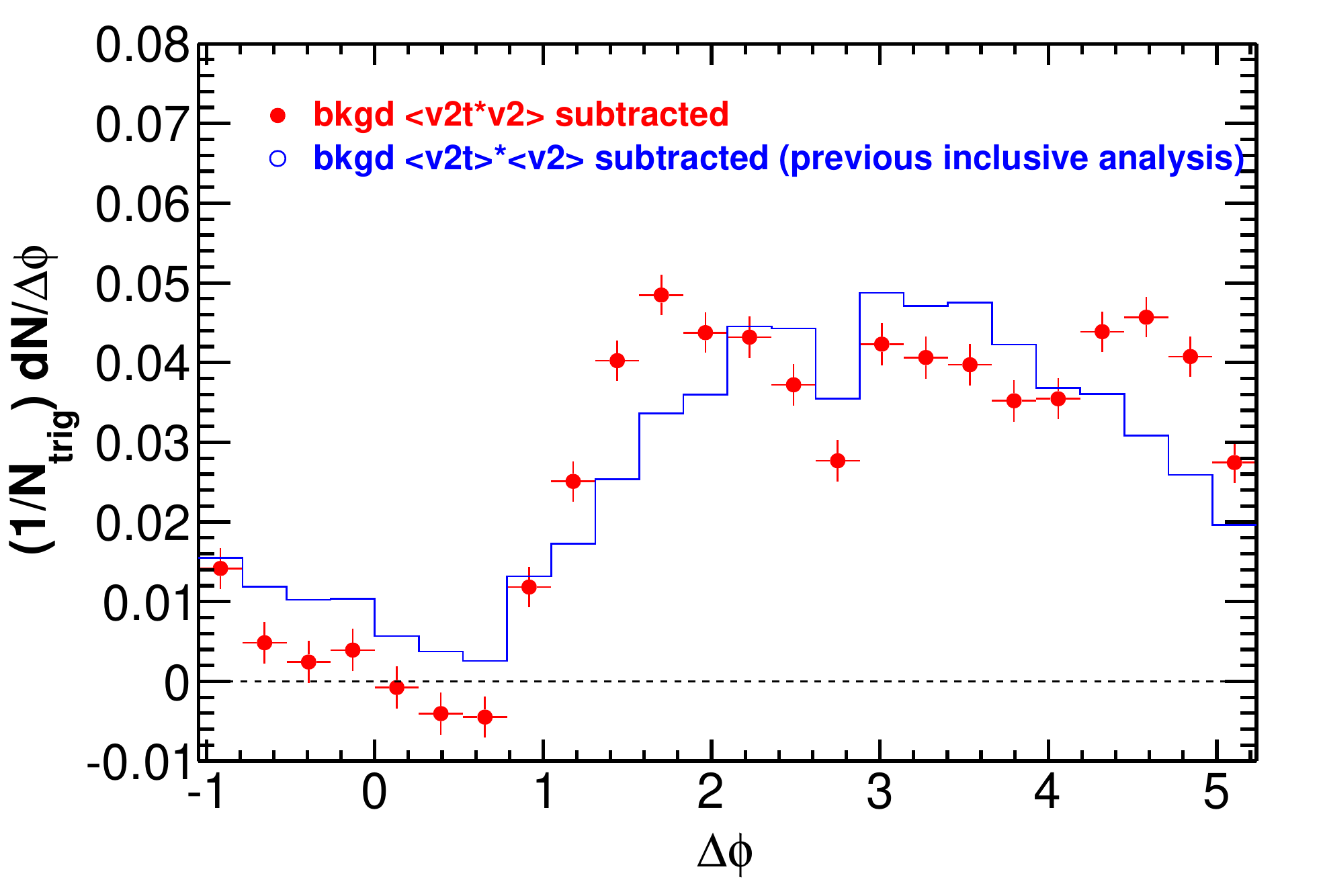}
  \end{center}
  \caption{(Color online) Effect of possible $\phis$-dependent elliptic flow anisotropy on inclusive dihadron correlations. Left panel: raw $\dphi$ correlation together with flow background obtained by two different ways, one by the average of the product of the trigger $\vtR$ and the associated particle $\va(\phis)$ as from this analysis (red histogram), and the other by the product of the average trigger and associated $v_2$ as from the standard inclusive dihadron correlation analysis (blue histogram). Right panel: the correlation signals subtracted by the background from this analysis (red points) and by the standard background from inclusive dihadron correlation analysis (blue histogram). The data are from minimum-bias 20-60\% Au+Au collisions. The trigger and associated particle $\ppt$ ranges are $3<\ptt<4$~\gev\ and $1<\pta<1.5$~\gev, respectively. A $|\deta|>0.7$ cut is applied to the trigger-associated pairs. Error bars are statistical.}
  \label{fig:inclusive_dihadron}
\end{figure}

Figure~\ref{fig:inclusive_dihadron} illustrates the effect. The left panel shows the raw dihadron correlation for $3<\ptt<4$~\gev\ and $1<\pta<1.5$~\gev\ together with two flow background curves both \zyam-normalized. The blue histogram is from a traditional inclusive dihadron correlation analysis with the $v_2$ modulation calculated from $\mn{\fff{2}{t}{2}}\cdot\mn{\fff{2}{a}{2}}$. 
The red histogram is that calculated from the $\phis$-dependent $\ff{2}{2}(\phis)$ by $\mn{\fff{2}{t}{2}(\phis)\cdot\fff{2}{a}{2}(\phis)}$ which is the correct flow background provided $v_2(\phis)$ is the real flow. 
(The $v_3$ and $v_4$ contributions are included in both flow background histograms). 
As seen from Fig.~\ref{fig:inclusive_dihadron}, the traditional flow background is underestimated. The right panel of Fig.~\ref{fig:inclusive_dihadron} shows the dihadron correlation signals after subtraction of the traditional background in the histogram and by the correct flow background in the data points, respectively. The signal from the traditional average flow background subtraction is less double-peaked. This means, if the ridge is entirely due to flow that must be $\phis$-dependent, then all the inclusive dihadron correlation analyses have under-subtracted the flow background, resulting in a more peaked away-side correlation signal.

\section{Effect of dipole fluctuations}
We have neglected the effect of dipole fluctuations (rapidity-even $v_1$) in flow background subtraction of Eq.~(\ref{eq}). Preliminary data~\cite{STAR_dipole} indicate that the dipole fluctuation effect changes sign at $\ppt\approx1$~\gev, negative at lower $\ppt$ and positive at higher $\ppt$. For $\pta=1$-2~\gev\ shown in Fig.~\ref{fig}, the dipole fluctuation effect is approximately zero and may be neglected. The qualitative conclusions on the near-side and away-side correlations are unaffected by the potential dipole fluctuations. 

\section{Effect of possible biases in event-plane reconstruction}
In our analysis, the event plane is reconstructed by particles excluding those within $|\deta|<0.5$ of the trigger particle. Question remains how much is the effect of possible biases to the reconstructed event plane by particles correlated to the trigger, especially on the away side. One way to estimate this possible effect is to analyze dihadron correlations relative to the event plane reconstructed from particles without excluding those within $|\deta|<0.5$ of the trigger, thereby maximizing the biases from jet-correlations. Our study shows that introducing a stronger bias in event-plane reconstruction causes a relatively small change in the correlation signals. This suggests that possible event-plane biases in our default results in Fig.~\ref{fig} may be also relatively small.

\section{Conclusions}
In this talk, we revisit our dihadron correlations relative to the event plane in arXiv:1010.0690v1. We focused on the $\dphi$ correlations at $|\deta|>0.7$ to study the ridge and away-side correlations. 
We subtract background from triangular anisotropy measured by the two-particle cumulant method with a $\eta$-gap ($\etagap$) of 0.7. The qualitative feature of the correlation data is unchanged. The ridge, with a reduced magnitude, is still present for in-plane trigger particles. It decreases from in-plane to out-of-plane. 

We consider the effect of a $v_n$ that is dependent on the trigger-particle $\phis$. We analyze the two-particle cumulants $\vpt{n}$ in events with different trigger particle $\phis$ separately. The second harmonic $\vpt{2}$ is found to decrease with increasing $\phis$. This is synonymous to the decreasing ridge magnitude with $\phis$. The fourth harmonic $\vpt{4}$ is found to also depend on $\phis$, but its effect on dihadron correlation is negligible. The third harmonic $\vpt{3}$ is found to be independent of $\phis$. The dihadron correlations are studied relative to the event plane with the subtraction of the two-particle cumulant anisotropies, $\veta{3}{0.7}$, $\veta{4}{0.7}$ and the $\phis$-dependent $\vpteta{2}{0.7}(\phis)$. With this exploratory subtraction of the $v_n$ values, the ridge is found to be eliminated. However, this result does not enlighten the origin of the ridge because the measured $v_n$ have likely already included the ridge; whether the ridge is due to flow or nonflow is undetermined. 

With the wide range of flow subtraction, the away-side double-peak structure for out-of-plane triggers remains robust. This is true even with the subtraction of $\phis$-dependent $\vpteta{n}{0.7}$. This may indicate path-length dependent modifications to the away-side jet traversing the sQGP medium.

\end{document}